\documentclass[twocolumn,tighten,times]{aastex631}

\newcommand{\mcl}[1]{\textcolor{black}{#1}} 
\usepackage{amsmath}

\begin{document}

\title{Scaling of Particle Heating in Shocks and Magnetic Reconnection}

\author[0000-0003-2191-1025]{Mitsuo Oka}
\affiliation{Space Sciences Laboratory, University of California Berkeley, 7 Gauss Way, Berkeley, CA 94720, USA}

\author[0000-0002-6924-9408]{Tai D. Phan}
\affiliation{Space Sciences Laboratory, University of California Berkeley, 7 Gauss Way, Berkeley, CA 94720, USA}

\author[0000-0003-3112-1561]{Marit \O{}ieroset}
\affiliation{Space Sciences Laboratory, University of California Berkeley, 7 Gauss Way, Berkeley, CA 94720, USA}

\author[0000-0003-1304-4769]{Daniel J. Gershman}
\affiliation{NASA Goddard Space Flight Center, Greenbelt, MD, USA}

\author[0000-0001-7188-8690]{Roy B. Torbert}
\affiliation{Physics Department, University of New Hampshire, Durham, NH, USA}
\affiliation{Southwest Research Institute, San Antonio, TX, USA}

\author[0000-0003-0452-8403]{James L. Burch}
\affiliation{Southwest Research Institute, San Antonio, TX, USA}

\author[0000-0001-7024-1561]{Vassilis Angelopoulos}
\affiliation{Earth, Planetary, and Space Sciences Department, and Institute of Geophysics and Planetary Physics, University of California, Los Angeles, Los Angeles, CA 90095, USA}










\begin{abstract}
Particles are heated efficiently through energy conversion processes such as shocks and magnetic reconnection in collisionless plasma environments. While empirical scaling laws for the temperature increase have been obtained, the precise mechanism of energy partition between ions and electrons remains unclear. Here we show, based on coupled theoretical and observational scaling analyses, that the temperature increase, $\Delta T$, depends linearly on three factors: the available magnetic energy per particle, the Alfv\'{e}n Mach number (or reconnection rate),  and the characteristic spatial scale $L$. Based on statistical datasets obtained from Earth's plasma environment, we find that $L$ is on the order of (1) the ion gyro-radius for ion heating at shocks, (2) the ion inertial length for ion heating in magnetic reconnection, and (3) the hybrid inertial length for electron heating in both shocks and magnetic reconnection. With these scales, we derive the ion-to-electron ratios of temperature increase as $\Delta T_{\rm i}/\Delta T_{\rm e} = (3\beta_{\rm i}/2)^{1/2}(m_{\rm i}/m_{\rm e})^{1/4}$ for shocks and $\Delta T_{\rm i}/\Delta T_{\rm e} = (m_{\rm i}/m_{\rm e})^{1/4}$ for magnetic reconnection, where $\beta_{\rm i}$ is the ion plasma beta, and $m_{\rm i}$ and $ m_{\rm e}$ are the ion and electron particle masses, respectively. We anticipate that this study will serve as a starting point for a better understanding of particle heating in space plasmas, enabling more sophisticated modeling of its scaling and universality.
\end{abstract}



\section{Introduction} \label{sec:intro}

\begin{deluxetable*}{lLlLlll}[t]
\tablecaption{Empirically derived scaling of particle heating during reconnection. \label{tab}}
\tablehead{\colhead{Species} & \colhead{Relation} & \colhead{Region} & \colhead{Plasma beta} & \colhead{Alfv\'{e}n speed $V_{\rm A}$} & \colhead{Mission} & \colhead{Reference}}
\startdata
ions & \Delta T_i \sim 0.13\,\,\, m_{\rm i}V_{\rm A}^2     & magnetopause & \beta_i \sim 0.2 - 550 & $< 600$ km/s & {\it THEMIS} & \cite{PhanTD_2014_Ion} \\
ions & \Delta T_i \lesssim 0.13\,\,\, m_{\rm i}V_{\rm A}^2 & magnetotail & \beta_i \sim 0.003 - 1 & $800 - 4000$ km/s & {\it MMS} & \cite{OierosetM_2024_Scaling} \\
electrons & \Delta T_e \sim 0.017\, m_{\rm i}V_{\rm A}^2        & magnetopause & \beta_e \sim 0.1 - 10 & $< 600$ km/s & {\it THEMIS} & \cite{PhanTD_2013_Electron} \\
electrons &\Delta T_e \sim 0.020\, m_{\rm i}V_{\rm A}^2        & magnetotail & \beta_e \sim 0.001 - 0.1 & $800 - 4000$ km/s & {\it MMS} & \cite{OierosetM_2023_Scaling} \\
\enddata
\tablecomments{The data were obtained by NASA's spacecraft missions, {\it Time History of Events and Macroscale Interactions during Substorms  (THEMIS)}  \citep{AngelopoulosV_2008_THEMIS} and {\it  Magnetospheric MultiScale (MMS)} \citep{BurchJL_2016_Magnetospheric}.}
\end{deluxetable*}

Particles -- both ions and electrons -- are heated efficiently through fundamental plasma  processes such as shocks and magnetic reconnection in space, solar, and astrophysical plasma environments  \citep[e.g.][]{TidmanDA_1971_Shock, TsurutaniBT_1985_Collisionless,  LinY_1994_Structure, KivelsonMG_1995_Introduction, PriestE_2000_Magnetic, AschwandenMJ_2005_Physics,  ZweibelEG_2009_Magnetic, LongairMS_2011_High, EastwoodJP_2013_Energy}.  A challenge is that, at magnetohydro-dynamic (MHD) scales, the conservation laws can only predict the {\it total} temperature on the downstream side. It remains unclear how the upstream energy is partitioned between ions and electrons during the energy conversion processes such as shocks and magnetic reconnection, despite extensive studies with observations and simulations as summarized below.

For non-relativistic collisionless shocks, the Rankine-Hugoniot (RH) relations reveal that, for the strong shock limit with the shock compression ratio of 4,  the increase of the total temperature  is approximated as $\Delta T \sim (3/16)m_{\rm i}V_{\rm in}^2$, where $m_{\rm i}$ is the proton mass and $V_{\rm in}$ is the bulk flow speed of the upstream, incoming plasma in the shock-rest, shock normal incidence frame \citep[e.g.][]{TidmanDA_1971_Shock, GhavamianP_2013_ElectronIon}. Because of the large ion-to-electron mass ratio, this total temperature change is comparable to the ion temperature change, $\Delta T \sim \Delta T_{\rm i}$. For electron heating, previous studies have identified that the increase of the electron temperature $\Delta T_{\rm e}$  correlates with the change of bulk flow energy, as well as the shock potential \citep[e.g.][]{ThomsenMF_1987_Strong, SchwartzSJ_1988_Electron, HullAJ_2000_Electron}. For example, it was reported that  $\Delta T_{\rm e} \sim 0.057\, \Delta \left(m_{\rm i}V_{n}^2/2\right)$, where $V_n$ is the bulk flow velocity along the shock normal direction $\mathbf{\hat{n}}$ and  $\Delta(m_{\rm i}V_{n}^2/2)$ is the change of bulk flow energy across the shock front \citep{HullAJ_2000_Electron}. Interestingly, a compilation of statistical data from Earth's and Saturn's bow shock shows that the temperature ratio $T_{\rm e}/T_{\rm i}$ on the downstream side decreases from $\sim$1 to $\sim$0.1 as the Alfv\'{e}n or magnetosonic Mach number on the upstream side increases from $\sim$1 to $\sim$10 \citep{GhavamianP_2013_ElectronIon}. To interpret this observational relation, a thermodynamic relation  for the minimum values of $T_{\rm e}/T_{\rm i}$ have been derived \citep{VinkJ_2015_electronion}, and various particle simulations have been conducted \cite[e.g.][]{TranA_2020_Electron, BohdanA_2020_Kinetic}. A possible extension to astrophysical parameter ranges is also important \citep[e.g.][]{GhavamianP_2013_ElectronIon, RaymondJC_2023_Electron}.

For magnetic reconnection,  {\it in-situ} observations in Earth's magnetosphere have shown that $\Delta T$ depends roughly linearly on available magnetic energy per particle $m_{\rm i}V_{\rm A}^2$, as summarized in Table \ref{tab}, where $V_{\rm A}$ is the Alfv\'{e}n speed measured in the upstream inflow region with inflow density and inflow magnetic field strength  \citep{PhanTD_2013_Electron,PhanTD_2014_Ion,OierosetM_2023_Scaling,OierosetM_2024_Scaling}. It is evident that the scaling applies over  large ranges of plasma $\beta$ and $V_{\rm A}$. Theories indicate that a similar linear dependence $\Delta T_{\rm i} = (1/3)m_{\rm i}V_{\rm A}^2$ arises from Alfv\'{e}nic counter-streaming proton beams in the exhaust \citep{DrakeJF_2009_Ion} and that the difference from the observations in the factor for the scaling can be attributed to the fact that the observed outflow jets are not always being fully Alfv\'{e}nic \citep{HaggertyCC_2015_competition, HaggertyCC_2018_reduction} due to the presence of electric potentials in the reconnection exhaust \citep[e.g.][]{EgedalJ_2005_situ, EgedalJ_2013_review}. For electron heating, a linear dependence of $\Delta T_{\rm e} \sim 0.03 - 0.05 m_{\rm i}V_{\rm A}^2$ has been obtained by particle simulations \citep{ShayMA_2014_Electron, HaggertyCC_2015_competition}. 

In this \mcl{Paper}, we propose an alternative approach to interpreting and predicting particle heating at shocks and in reconnecting current sheets. We propose that heating is a function of the available magnetic energy per particle, the Alfv\'{e}n Mach number, and the characteristic spatial scale. \mcl{Heating is defined as an increase in temperature, which corresponds to the second-order moment of the velocity distribution in the plasma rest frame.} The paper is organized as follows. In Section \ref{sec:model}, we deduce and normalize the expression. Then, the validity of the idea is tested against observations of quasi-perpendicular shocks (Section \ref{sec:shock}) and magnetic reconnection (Section \ref{sec:mrx}).  Finally, we summarize and discuss the results in Section \ref{sec:discuss}.

\section{Theoretical expression for particle heating}\label{sec:model}

In general, an energization $\Delta \varepsilon$ of a charged particle can be expressed as  $\Delta \varepsilon = qVBL$ where $q$ is the particle charge, $VB$ is the motional electric field constructed from the characteristic flow speed $V$ and magnetic field $B$, and $L$ is the distance of travel during energization. \mcl{This is a ballpark estimate with an assumption the bulk flow $\mathbf{V}$ and the magnetic field $\mathbf{B}$ are perpendicular to each other.} A similar expression has been used in the context of particle acceleration to non-thermal energies \citep[e.g.][]{HillasM_1984_Origin, MatthaeusWH_1984_Particle, OkaM_2025_Maximum} and the rate of explosive energy release in Earth's magnetotail \citep{OkaM_2022_Electron}.

To better illustrate the physical processes involved in energization, we further rewrite this expression by normalizing $V$ and $L$ by $V_{\rm A}$ and the ion inertial length $d_{\rm i}$, respectively. We also convert kinetic energy $ \varepsilon$ to temperature $T$ via $T = (2/f)\varepsilon$ where $f$ is the degrees of freedom and is chosen to be $f=3$ throughout this paper. Then, we obtain
\begin{equation}\label{eq:Hillas}
    \Delta T = \frac{2}{f}\left(\frac{V}{V_{\rm A}}\right)\left(\frac{L}{d_{\rm i}}\right) m_{\rm i}V_{\rm A}^2
\end{equation}
Therefore, any scaling for heating should be explained by three  physical parameters: (1) the degrees of freedom $f$, (2) the normalized flow speed $V/V_{\rm A}$, typically referred to as the Alfv\'{e}n Mach number $M_{\rm A}$ or reconnection rate $\alpha_{\rm R}$, and (3) the characteristic spatial scale $L/d_{\rm i}$. The quantity $m_{\rm i}V_{\rm A}^2$ is an environmental parameter that depends on the density $N$ and magnetic field strength $B$. It is derived by dividing the incoming Poynting flux by the inflowing particle density flux \citep[e.g.][]{ShayMA_2014_Electron} and therefore represents the magnetic energy per particle that is available in the upstream region of a shock or a reconnecting current sheet. It can also be used to normalize the temperature $\Delta T$. 

All these parameters, except $L$, are based on upstream values and are also observable, although there may be some uncertainty in the measurement of the reconnection rate $\alpha_{\rm R}$.  The parameter $L$ is not directly known, but various characteristic lengths, such as the gyro-radius and inertial length, can be considered to calculate the predicted heating from Eq.(\ref{eq:formula-shock}) and compare with observational data to deduce the most appropriate  scale for $L$. As we shall see in the following section, the values of $L$ are typically on the order of the ion kinetic scale for ion heating and the hybrid inertial length for electron heating.  It is also worth noting that $L/d_{\rm i}$ can be replaced with $\Omega_{\rm ci}\tau_{\rm A}$, where $\Omega_{\rm ci}$ is the ion cyclotron frequency and $\tau_{\rm A}$ is the Alfv\'{e}n transit time. This alternative formulation is useful when interpreting the time evolution of particle energization \citep[e.g.][]{MatthaeusWH_1984_Particle}.

\section{Application of the formula to shocks}\label{sec:shock}

We first apply Eq.(\ref{eq:Hillas}) to the physics of non-relativistic collisionless shocks. Using $f=3$ and $M_{\rm A}=V_{\rm in}/V_{\rm A}$, where $V_{\rm in}$ is the incoming flow speed in the shock normal incidence frame, one obtains
\begin{equation}\label{eq:formula-shock}
    \Delta T = \frac{2}{3}M_{\rm A}\left(\frac{L}{d_{\rm i}}\right) m_{\rm i}V_{\rm A}^2
\end{equation}
As noted in the previous section, all quantities except $L$ are observable. Therefore, we use observational datasets to find  reasonable values for $L$. Specifically, we analyzed \mcl{80} cases of supercritical, quasi-perpendicular bow shock crossings obtained by the {\it Magnetospheric MultiScale (MMS)} mission. 

These events were selected through visual inspection, with upstream and downstream time periods defined to calculate the temperature increases for both ions ($\Delta T_{\rm i}$) and electrons ($\Delta T_{\rm e}$). We focused on super-critical shocks because they are much more frequently observed. Also, we avoided the events that are confounded by upstream-escaping particles, resulting in a bias toward quasi-perpendicular shocks. 

\begin{figure*}
\plotone{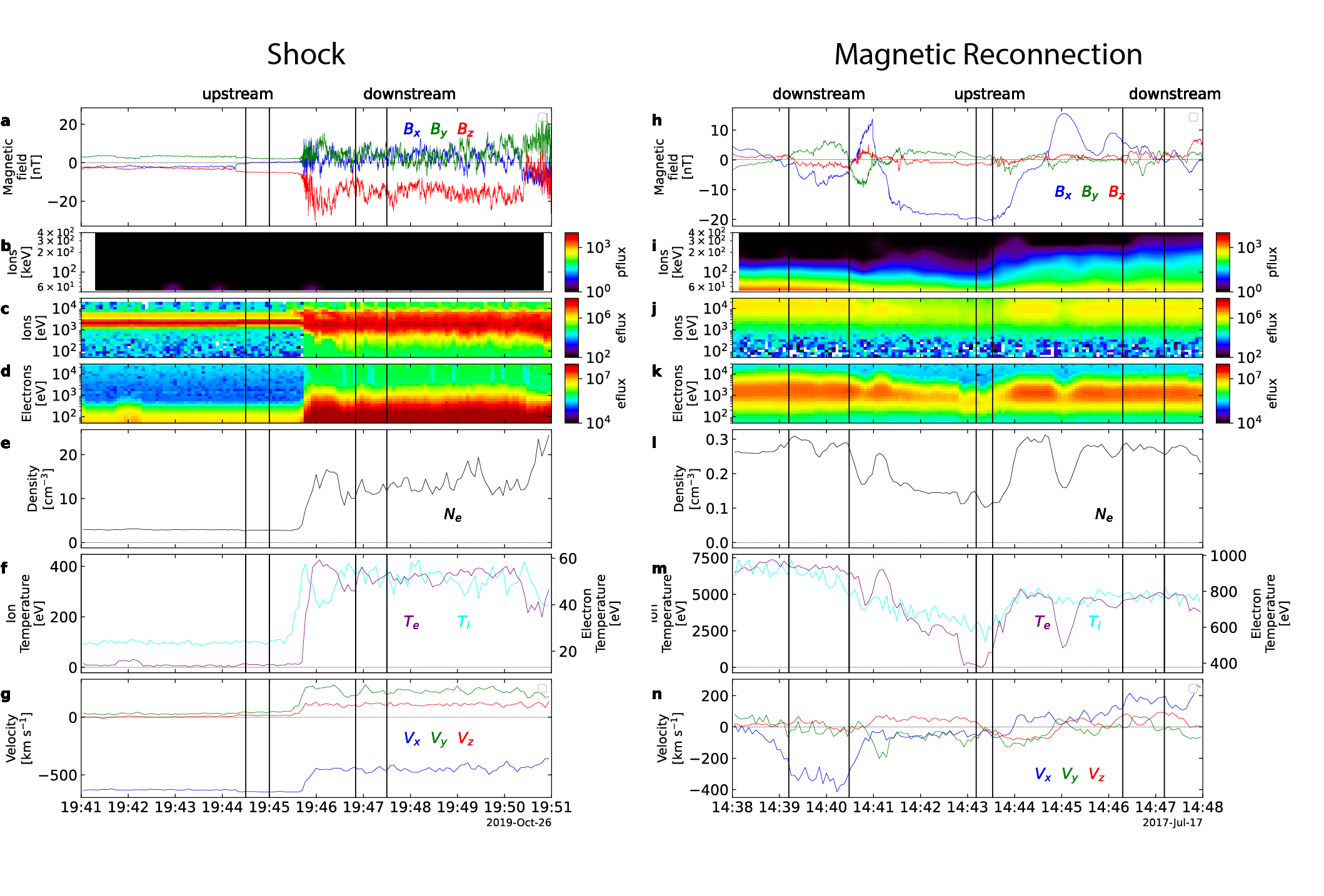}
\caption{\mcl{Plasma parameters of example shock (left) and reconnection (right) events, demonstrating our procedure of event identification and analysis. From top to bottom are (a,h) magnetic field vector, (b,i) energetic ion spectrogram, (c,j) ion spectrogram, (d,k) electron spectrogram, (e, l) electron density, (f, m) ion and electron temperatures, (g, n) ion bulk flow vector. For the color scale in spectrograms, pflux and eflux represent particle flux (cm$^{-2}$s$^{-1}$sr$^{-1}$keV$^{-1}$) and energy flux (keV cm$^{-2}$s$^{-1}$sr$^{-1}$keV$^{-1}$), respectively.}\label{fig:event_overview}}
\end{figure*}

\mcl{While more details of our event selection and analysis procedures are described in Appendix, we present here an example crossing event, shown in Figure \ref{fig:event_overview} (left), observed on  2019 October 26. In this event, {\it MMS} traversed from upstream to downstream and detected a significant increase in ion and electron temperatures (Fig.\ref{fig:event_overview}f) accompanied by similarly abrupt changes in the magnetic field (Fig.\ref{fig:event_overview}a), density (Fig.\ref{fig:event_overview}e), and flow velocity (Fig.\ref{fig:event_overview}g). We then selected stable intervals, highlighted by the vertical lines, to determine upstream and downstream plasma parameters, especially the temperatures.  The upstream interval was carefully chosen between the magnetic field direction change at 19:44:30 and the leading edge of the region of reflected-gyrating ions (Fig.\ref{fig:event_overview}c). We repeated this procedure for all events and performed a statistical analysis, as described below.}

\begin{figure*}
\plotone{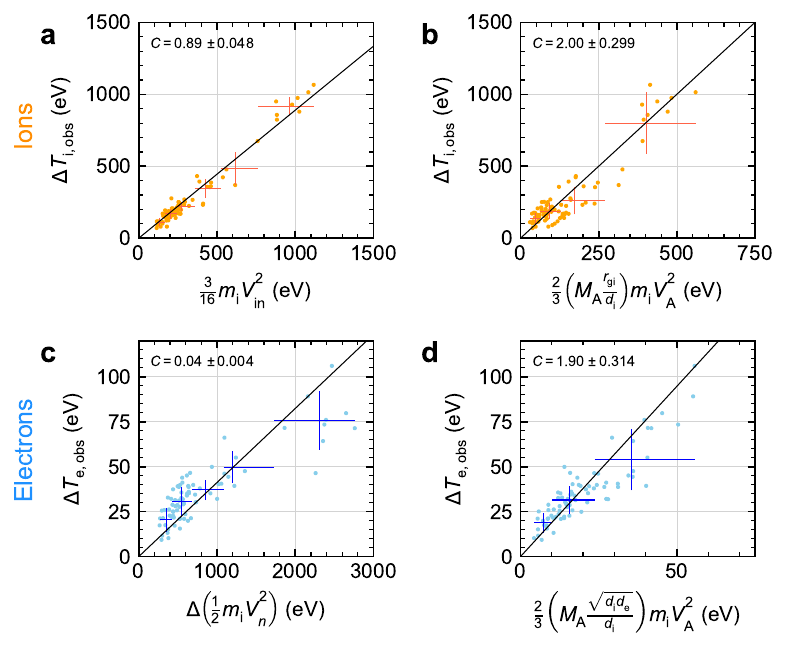}
\caption{Scaling of the temperature increase across Earth's bow shock for ions (upper panels) and electrons (lower panels). A new approach using Eq.(\ref{eq:formula-shock}) (right panels) is compared to an earlier approach of using the bulk flow energy (left panels). The titles of the horizontal axes indicate the formula used to predict the temperature increase. \mcl{The multiplication factor $C$ is derived through linear regression and illustrated by the thin black line.}\label{fig:shock}}
\end{figure*}

Figure \ref{fig:shock} shows the results of our \mcl{statistical} analysis for both ions (upper panels) and electrons (lower panels). 
For ions, we first examined the validity of the classical scaling \mcl{$\Delta T \sim \Delta T_{\rm i}\sim(3/16)m_{\rm i}V_{\rm in}^2$. To this end, we have introduced a multiplicative factor $C$ to assume $\Delta T_{\rm i} = C(3/16)m_{\rm i}V_{\rm in}^2$ and performed a linear regression to derive the best-fit value of $C$.  For the linear regression, we used} the average values and standard deviations calculated for each dataset, separated into bins. Figure \ref{fig:shock}a shows the result.  It is evident that the classical scaling works with the best-fit value of $C \sim 0.88$, close to but somewhat less than the expected value of 1.  

\mcl{We then tested Eq.(\ref{eq:formula-shock}) more directly by comparing with observations with different choices of $L$, such as the ion inertial length $d_{\rm i}$ and the gyro-radius based on the upstream ion thermal speed $r_{\rm gi}$.  More precisely, we again introduced a multiplicative factor $C$ and derived the best-fit value of $C$ for $L=Cd_{\rm i}$ and $L=Cr_{\rm gi}$. We obtained $L \sim 2.5d_{\rm i}$ with c.c. = 0.79, and $L \sim 2.00r_{\rm gi}$ with c.c.=0.93, where c.c. is the Pearson correlation coefficient. Therefore, we conclude that Eq.(\ref{eq:formula-shock}) with $L=2r_{gi}$ gives a reasonable, semi-empirically-derived formula for ion heating at shocks, and this is shown in Figure \ref{fig:shock}b. We could also try the ion gyro-radius based on the incoming flow speed $r_{\rm gi0} = V_{\rm in}/\Omega_{\rm ci}$  and the foot size $L_{\rm foot}$ as formulated by \cite{LiveseyWA_1984_comparison}. However, with these parameters, Eq.(\ref{eq:formula-shock}) reduces to $\Delta T_{\rm i} \propto m_{\rm i}V_{\rm in}^2$, which is essentially the same as the classical scaling of $\Delta T_{\rm i}\sim(3/16)m_{\rm i}V_{\rm in}^2$. In fact, this classical scaling can be obtained by inserting $L = (9/32)r_{\rm gi0} \sim (1/4)r_{\rm gi0}$ into Eq.(\ref{eq:formula-shock}).}


If we further introduce the ion plasma beta $\beta_{\rm i}$, the formula can be rewritten as
\begin{equation}\label{eq:shock-dTi}
\Delta T_{\rm i} = \frac{4}{3}M_{\rm A}\left(\frac{3}{2}\beta_{\rm i}\right)^{\frac{1}{2}}m_{\rm i}V_{\rm A}^2
\end{equation}
where we used the relation $r_{\rm gi}/d_{\rm i} = v_{\rm thi}/V_{\rm A} = (3\beta_{\rm i}/2)^{1/2}$. Here, $v_{\rm thi}$ is the upstream ion thermal speed defined as the root-mean-square speed of the three-dimensional Maxwellian velocity distribution, i.e., $v_{\rm thi}=\sqrt{3k_BT/m}$, where $k_B$ is the Boltzman constant.

For electrons (lower panels), we again started from validating earlier results. Figure \ref{fig:shock}c tests an earlier idea that  $\Delta T_{\rm e}$  depends on the change of bulk flow energy $\Delta (m_{\rm i}V_n^2/2)$ across the shock front \citep[e.g.][]{SchwartzSJ_1988_Electron, HullAJ_2000_Electron}. We found that $\Delta T_{\rm e}$ increases linearly with $\Delta (m_{\rm i}V_n^2/2)$ \mcl{, with a slope of $C \sim 0.04$, which is comparable to but somewhat smaller than $\sim$0.06 reported by \cite{HullAJ_2000_Electron}. This difference appears to be due to the fact that, in our cases, the observed values of $\Delta T_{\rm e}$ increased progressively less as $\Delta (m_{\rm i}V_n^2/2)$ increased.}

\mcl{For the application of Eq.(\ref{eq:formula-shock}), we tested a variety of different parameters such as electron inertial length $d_{\rm e}$, electron gyro-radius $r_{\rm ge}$, the hybrid inertial length $\sqrt{d_{\rm i}d_{\rm e}}$ and hybrid gyro-radius $\sqrt{r_{\rm gi}r_{\rm ge}}$. With the linear regression analysis, we obtained $L=(12.4\pm2.1)d_{\rm e}$ with c.c. $=0.90$ and $L=(9.8\pm1.6)r_{\rm e}$ with c.c.$=0.88$, indicating that the electron scale underestimates the spatial scale and hence electron heating by an order of magnitude. In contrast, the hybrid scale provides a better estimate as we obtained $L=(1.90\pm0.31)\sqrt{d_{\rm i}d_{\rm e}}$ with c.c. $=0.90$ and $L=(1.55\pm0.26)\sqrt{r_{\rm gi}r_{\rm ge}}$ with c.c. $=0.87$.  While $\sqrt{d_{\rm i}d_{\rm e}}$ and $\sqrt{r_{\rm gi}r_{\rm ge}}$ are comparable, we adopt $\sqrt{d_{\rm i}d_{\rm e}}$ because it produces a slightly higher c.c., and the formula remains simple. Then, within the error range, electron heating at shocks can be reasonably modeled by Eq.(\ref{eq:formula-shock}) with $L = 2\sqrt{d_{\rm i}d_{\rm e}}$. Figure \ref{fig:shock}d shows the regression analysis for the choice of $\sqrt{d_{\rm i}d_{\rm e}}$. } Because of the relation $\sqrt{d_{\rm i}d_{\rm e}}/d_{\rm i} = \sqrt{d_{\rm e}/d_{\rm i}} = (m_{\rm i}/m_{\rm e})^{-1/4}$, the formula can be expressed as
\begin{equation}\label{eq:shock-dTe}
\Delta T_{\rm e} = \frac{4}{3}M_{\rm A}\left(\frac{m_{\rm i}}{m_{\rm e}}\right)^{-\frac{1}{4}}m_{\rm i}V_{\rm A}^2
\end{equation}

\begin{figure*}
\plotone{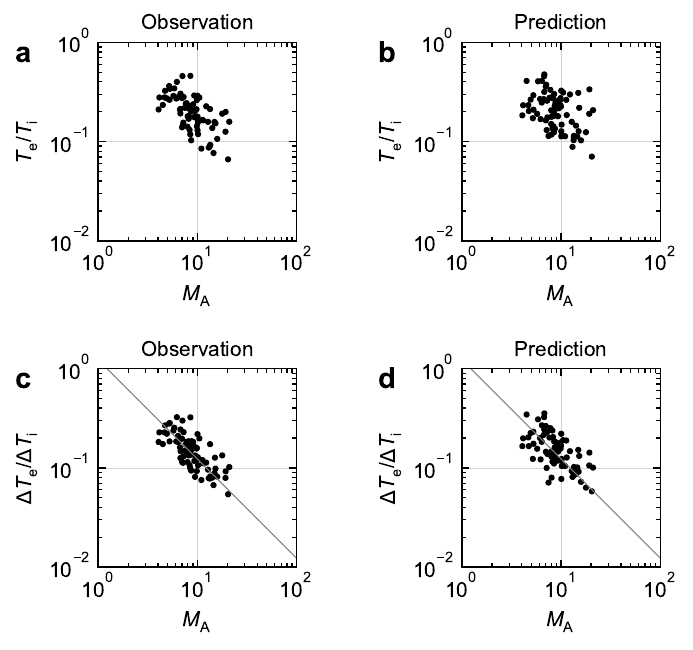}
\caption{The $M_{\rm A}$ dependence of the electron-to-ion ratios of temperatures (upper panels) and temperature increases (lower panels), obtained from observations at Earth's bow shock (left panels) and  predictions based on Eq.(\ref{eq:formula-shock}) (right panels). While each observational data point (filled circle) corresponds to a different value of $\psi$, the relation expressed by Eq.(\ref{eq:shock_Tratio-Ma}) for $\psi=8$ is shown by the thin gray curve.
\label{fig:shock-Tratio}}
\end{figure*}

\mcl{By using the derived formula for ions and electrons}, we can examine the electron-to-ion ratio of downstream temperature ($T_{\rm e}/T_{\rm i}$) using both observations and predictions based on Eq.(\ref{eq:formula-shock}), as shown in the upper panels of Figure \ref{fig:shock-Tratio}. Figure \ref{fig:shock-Tratio}a is obtained from our observational {\it MMS} dataset and is consistent with \cite{GhavamianP_2013_ElectronIon} who reported that $T_{\rm e}/T_{\rm i}$  on the downstream side decreases from $\sim$1 to $\sim$0.1 as the upstream Alfv\'{e}n Mach number  increases from $\sim$1 to $\sim$10. Figure \ref{fig:shock-Tratio}b shows the prediction based on our formulas with $L=2r_{\rm gi}$ for ions and $L=2\sqrt{d_{\rm i}d_{\rm e}}$ for electrons. When compared to the observation (Figure \ref{fig:shock-Tratio}a), it is evident that our simple formula reproduces the empirical scaling well and that $T_{\rm e}/T_{\rm i}$ correlates with $M_{\rm A}$. 

Figures \ref{fig:shock-Tratio}(c,d) show the $M_{\rm A}$ dependencies for $\Delta T_{\rm e}/\Delta T_{\rm i}$ instead of $T_{\rm e}/T_{\rm i}$. Again, a similar linear relation was obtained for both observations (Figure \ref{fig:shock-Tratio}c) and  predictions (Figure \ref{fig:shock-Tratio}d). The origin of this $M_{\rm A}$-dependence can be understood when we combine the formulas, i.e., Eq.(\ref{eq:formula-shock}) with $L=2r_{\rm gi}$ for $\Delta T_{\rm i}$ and $L=2\sqrt{d_{\rm i}d_{\rm e}}$ for $\Delta T_{\rm e}$, to obtain
\begin{align}
    \label{eq:shock-TeTi}
    \frac{\Delta T_{\rm e}}{\Delta T_{\rm i}} &= \frac{\sqrt{d_{\rm i}d_{\rm e}}}{r_{\rm gi}} \\
    &= \frac{V_{\rm A}}{v_{\rm thi}}\left(\frac{m_{\rm i}}{m_{\rm e}}\right)^{-\frac{1}{4}} \label{eq:shock_Tratio-vthi}\\
    &= \psi \left(\frac{m_{\rm i}}{m_{\rm e}}\right)^{-\frac{1}{4}}M_{\rm A}^{-1} \label{eq:shock_Tratio-Ma}
\end{align}
where we introduced $\psi=V_{\rm in}/V_{\rm thi}$. Therefore, the empirically known linear relationship between  $T_{\rm e}/T_{\rm i}$  (or equivalently $\Delta T_{\rm e}/\Delta T_{\rm i}$) and $M_{\rm A}$ is attributed to the characteristic of the solar wind, in which $\psi$  vary less than $M_{\rm A}$. In our dataset, $\psi$ ranges from $\sim$3 to $\sim$20, with a typical value of $\sim$8. The gray lines in Figure \ref{fig:shock-Tratio}(c,d) show the prediction from Eq.(\ref{eq:shock_Tratio-Ma}) with a fixed value of $\psi=8$. The predicted data points (filled circles) in Figure \ref{fig:shock-Tratio} are based on observed values of $\psi$, which vary from $\sim3$ to $\sim 20$.

Furthermore, the above expressions, in particular Eq.(\ref{eq:shock_Tratio-vthi}), or a direct comparison between Eqs. (\ref{eq:shock-dTi}) and (\ref{eq:shock-dTe}) reveal that the previously-known $M_A$ dependence is intrinsically a $\beta_{\rm i}$ dependence. In fact,   $\beta_{\rm i} = (2/3)(v_{\rm thi}/V_{\rm A})^2 = (2/3)\psi^{-2} M_{\rm A}^2$ and this relation in the upstream parameters can be confirmed with our dataset (Appendix). Consequently, we arrive at
\begin{equation}\label{eq:shock_Tratio-betai}
    \frac{\Delta T_{\rm i}}{\Delta T_{\rm e}} = \left(\frac{3}{2}\beta_{\rm i}\right)^{\frac{1}{2}}\left(\frac{m_{\rm i}}{m_{\rm e}}\right)^{\frac{1}{4}}
\end{equation}
where we deliberately considered $\Delta T_{\rm i}/\Delta T_{\rm e}$  instead of $\Delta T_{\rm e}/\Delta T_{\rm i}$ to avoid using negative indices and to more easily compare with the  same ratio of temperature increase for magnetic reconnection derived in the next section. 

\begin{figure}
\plotone{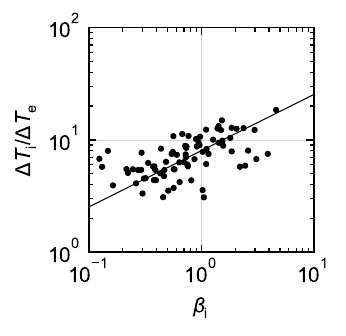}
\caption{The $\beta_{\rm i}$ dependence of the electron-to-ion ratio of temperature increase. \mcl{The observations at Earth's bow shock are shown by the filled circles, whereas the derived formula of $\Delta T_{\rm i}/\Delta T_{\rm e} = (3\beta_{\rm i}/2)^{1/2}(m_{\rm i}/m_{\rm e})^{1/4}$ is indicated by the black line. }
\label{fig:shock-beta}}
\end{figure}

Figure \ref{fig:shock-beta} shows such a $\beta_i$ dependence of $\Delta T_{\rm i}/\Delta T_{\rm e}$ obtained from both observations (filled circles) and the above expression (gray line). It is evident that the observation matches well with the \mcl{derived formula}.

\section{Application of the formula to magnetic reconnection}\label{sec:mrx}

Let us now consider how the expression (\ref{eq:Hillas}) applies to magnetic reconnection. We use the normalized reconnection rate $\alpha_{\rm R} =V_{\rm in}/V_{\rm A}$ where $V_{\rm in}$ is the reconnection inflow speed, so that Eq.(\ref{eq:Hillas}) becomes
\begin{equation}\label{eq:formula-mrx}
    \Delta T = \frac{2}{3}\alpha_{\rm R}\left(\frac{L}{d_{\rm i}}\right) m_{\rm i}V_{\rm A}^2
\end{equation}
Unfortunately, unlike the observations of Earth's bow shock where $M_{\rm A}$ can be measured directly, it is challenging to measure $\alpha_{\rm R}$ in Earth's magnetosphere because the plasma density becomes very low in the inflow region. Nevertheless, recent careful analysis with different methodologies reported $\alpha_{\rm R}$ values in a range of $0.15 - 0.20$ \citep[e.g.][]{NakamuraTKM_2018_Measurement, GenestretiKJ_2018_How}. \mcl{In fact,} a recent theoretical study argued that $\alpha_{\rm R}$ is a function of the opening angle made by the upstream magnetic fields and that it could be as high as $\alpha_{\rm R} \sim$ 0.2 \citep{LiuYH_2017_Why}. \mcl{On the other hand, } many simulation studies have established the canonical value of $\alpha_{\rm R} \sim 0.1$ \citep[e.g.][]{ShayMA_1999_scaling, BirnJ_2001_Geospace, LiuYH_2022_Firstprinciples}. Therefore, throughout this paper, we assume $\alpha_{\rm R}=0.1$ and explore different values of $L$. We shall keep in mind, however, that it is the product $\alpha_{\rm R}(L/d_{\rm i})$ that can be constrained by observations. It is very plausible that $\alpha_{\rm R}$ is $0.2$  and, in such a case, any values of $L$ we deduce in this paper should be reduced by half. 

\mcl{With this caveat in mind, we revisited previous statistical studies of plasma heating,  summarized in Table \ref{tab}. Figure \ref{fig:event_overview} (right) shows an example crossing of magnetotail reconnection observed by {\it MMS} on 2017 July 17 \citep[e.g.][]{OkaM_2022_Electron, OierosetM_2023_Scaling, OierosetM_2024_Scaling}. A flow reversal from tailward ($V_{\rm x}<0$) to earthward ($V_{\rm x}>0$) indicates that the X-line passed by the spacecraft in the plasma sheet with relatively small $B_{\rm x}$ (Fig.\ref{fig:event_overview}n). During the X-line passage, the spacecraft entered into a region of enhanced $B$ and depressed $N$, $T$, and $V$, indicating that the spacecraft moved into the lobe region temporarily. The periods of fast flows in the plasma sheet and the excursion to the lobe are highlighted by the vertical lines and marked as downstream and upstream respectively. The plasma temperatures averaged over these intervals are used to derive the temperature increase across reconnection. Essentially the same process was repeated for 20 magnetotail reconnection events reported by \cite{OierosetM_2023_Scaling, OierosetM_2024_Scaling} as well as 103 magnetopause reported by \cite{PhanTD_2013_Electron, PhanTD_2014_Ion}, although the magnetopause events are some what complicated due to the asymmetry in the inflow direction. See Appendix for more details and the full lists of events.}

\begin{figure*}
\plotone{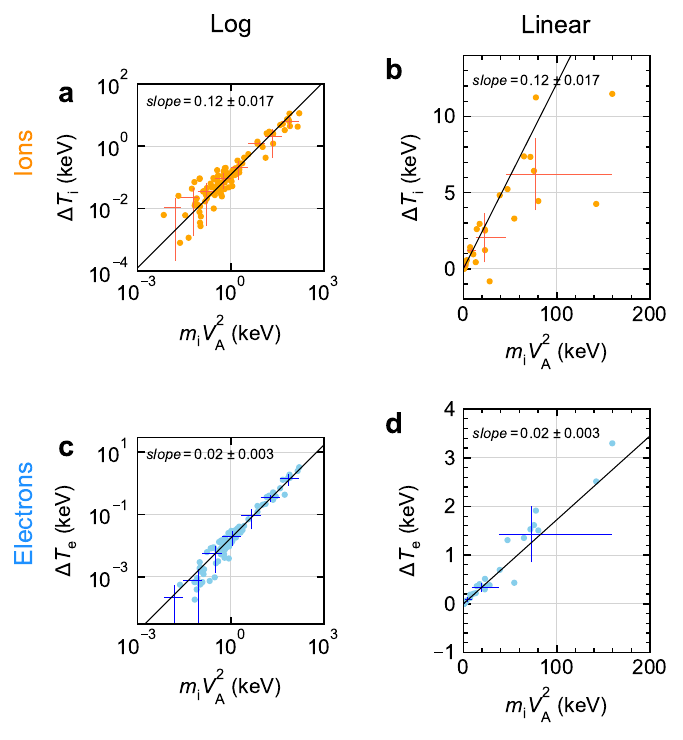}
\caption{Scaling of particle heating by magnetic reconnection, demonstrating \mcl{a linear dependence on $m_{\rm i}V_{\rm A}^2$ for  ions (upper panels) and electrons (lower panels).  A linear regression is performed to derive the slope, as annotated on the top left corner of each panel. The same data points and the linear regression results are shown in logarithmic (left panels) and linear (right panels) scales.} }
\label{fig:mrx}
\end{figure*}

\mcl{As in Section \ref{sec:shock}, we first attempted to reproduce earlier findings. Figure \ref{fig:mrx} shows the observed values of ion heating $\Delta T_{\rm i}$ (upper panels) and electron heating $\Delta T_{\rm e}$ (lower panels) as a function of $m_{\rm i}V_{\rm A}^2$ in logarithmic (left panels) and linear (right panels) scales. The data points from both magnetopause  and magnetotail events are combined so that they are spread over a wide range of values, as demonstrated in the logarithmic plots. The linear plots are shown because there were three negative values of $\Delta T_{\rm i}$ and six negative values of $\Delta T_{\rm e}$. The linear format also enhances the deviation of $\Delta T_{\rm i}$ from the overall trend for larger values of $m_{\rm i}V_{\rm A}^2$, as already reported by \cite{OierosetM_2023_Scaling}. The slopes obtained through linear regression of all data points including negative values are annotated at the top left corner of each panel, and they are consistent with the reported values (Table \ref{tab}) within the error ranges.}

Then, if we choose $L = 2d_{\rm i}$ for ion heating and insert this into Eq.(\ref{eq:formula-mrx}), we immediately obtain
\begin{equation}
\Delta T_i = \frac{2}{15}\,m_{\rm i}V_{\rm A}^2 \sim 0.13\, m_{\rm i}V_{\rm A}^2
\end{equation}
which is identical to the ion heating scaling of $0.13\,m_{\rm i}V_{\rm A}^2$ obtained by \cite{PhanTD_2014_Ion}, \cite{OierosetM_2024_Scaling}\mcl{, and our combined analysis described above.} Also, if we introduce a hybrid scale of $L = 2\sqrt{d_{\rm i} d_e}$ for electron heating, we obtain
\begin{equation}
\Delta T_e = \frac{2}{15} \left(\frac{m_{\rm i}}{m_{\rm e}}\right)^{-\frac{1}{4}} m_{\rm i}V_{\rm A}^2 \sim 0.020\, m_{\rm i}V_{\rm A}^2
\end{equation}
This is exactly the same as the scaling of $\sim 0.020\, m_{\rm i}V_{\rm A}^2$ obtained by \cite{OierosetM_2023_Scaling} and \mcl{our combined analysis described above, and also } consistent with $\sim 0.017\, m_{\rm i}V_{\rm A}^2$ reported by \cite{PhanTD_2013_Electron}. The value $L=2d_e$ would underestimate the electron heating because $d_e/d_{\rm i} \sim \sqrt{m_e/m_{\rm i}}$ and we obtain $\Delta T_e \sim 1.5\times10^{-3}\,m_{\rm i}V_{\rm A}^2$, an order of magnitude smaller than what we expect $\sim 0.02$. 

\mcl{Following our analysis in Section \ref{sec:shock}, we can again introduce a multiplicative parameter $C$ and hence $L=Cd_{\rm i}$ to obtain $\Delta T_{\rm i} = C(1/15)m_{\rm i}V_{\rm A}^2$ for ion heating. Using the statistically-derived slope, we immediately obtain $C=1.82 \pm 0.249$ with c.c.$=0.88$. Therefore, within the error range, we can reasonably justify the use of $L=2d_{\rm i}$ in the above formula. }

\begin{figure*}
\plotone{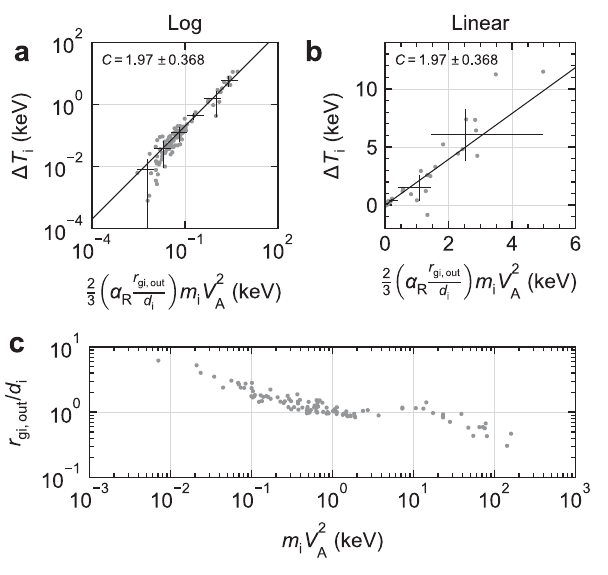}
\caption{\mcl{Additional analysis of ion temperature increase and ion gyro-radius during magnetic reconnection. (a) A comparison between the ion temperature increase and the formula based on Eq.(\ref{eq:formula-mrx}) with $L=Cr_{\rm gi,out}$ and $\alpha_R=0.1$, where $C$ is derived from a linear regression. (b) Same as (a) but displayed in linear scale.  (c) the ratio of $r_{\rm gi,out}/d_{\rm i}$ as a function $m_{\rm i}V_{\rm A}^2$, demonstrating that $r_{\rm gi,out}$ and $d_{\rm i}$ are often comparable to each other in Earth's magnetotail reconnection. }}
\label{fig:mrx-rgi-di}
\end{figure*}

\mcl{We also note} that $d_{\rm i}$ is often comparable to the ion gyro-radius $r_{\rm gi,out}$ in the plasma sheet, defined with the ion thermal speed in the \mcl{{\it outflow}} region. \mcl{This suggests} that $\Delta T_{\rm i}$ can be approximated with $L=2r_{\rm gi, out}$. Figure \ref{fig:mrx-rgi-di}(a,b) demonstrates that this alternative prediction with \mcl{$L=Cr_{\rm gi, out}$ yields a multiplicative factor of $C=1.97 \pm 0.37$ with c.c.$=0.95$, supporting this expectation.  Figure \ref{fig:mrx-rgi-di}c further demonstrates} that the ratio of $r_{\rm gi,out}/d_{\rm i}$ is \mcl{indeed} comparable to $\sim$1, especially when $m_{\rm i}V_{\rm A}^2$ is in the range of $0.1 - 10$. \mcl{However, this ratio deviates significantly when $m_{\rm i}V_{\rm A}^2$ is either below 0.1 or above 10. Moreover, in deriving our heating formula from a predictive standpoint, it is more reasonable to use $d_{\rm i}$, an upstream parameter, rather than $r_{\rm gi,out}$, a downstream parameter. Therefore, we retain the use of $d_{\rm i}$ in our formula.}

\mcl{For electron heating, we assume $L=C\sqrt{d_{\rm i}d_{\rm e}}$ to obtain $\Delta T_e = C(1/15)(m_{\rm i}/m_{\rm e})^{-1/4} m_{\rm i}V_{\rm A}^2$. The best-fit value $C$ that corresponds to the derived slope is $C = 1.739 \pm 0.258$  with c.c.$=0.98$. The error range allows for a maximum of $C = 1.998$, indicating that the derived value of $C$ is somewhat smaller than our assumed value of $C=2$ used in deriving the formula. This deviation likely originates from the magnetopause data, which suggests the relation of $\Delta T_{\rm e} \sim 0.017m_{\rm i}V_{\rm A}^2$ \citep{PhanTD_2013_Electron} instead of $\Delta T_{\rm e} \sim 0.020m_{\rm i}V_{\rm A}^2$. In fact, using the magnetotail data alone resulted in $C=1.96 \pm 0.362$. Given the complexity of analyzing magnetopause reconnection arising from asymmetry in the inflow direction, we retain our assumption of $C=2$ and defer more refined observational and theoretical studies to future work.  This is a reasonable choice, especially because, as we shall see below, the temperature ratio $\Delta T_{\rm i}/\Delta T_{\rm e}$ with the assumption of $C=2$ matches with the one derived from an independent theoretical study.  }

\mcl{To explore potential alternative assumptions, we also consider $L=C\sqrt{r_{\rm gi}r_{\rm ge}}$, which yields $C=1.50 \pm 0.379$ with c.c.  $=0.88$ and increases the complexity of the formula, similar to the case for shocks. While this alternative may be worth further investigation, its slightly lower agreement with the data and the increased complexity of the formula support our choice of $L=2\sqrt{d_{\rm i}d_{\rm e}}$ in the present study.}


\begin{figure}
\plotone{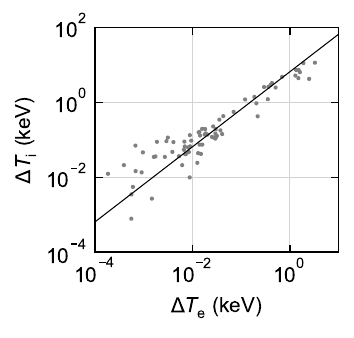}
\caption{Comparison between ion and electron temperature changes across magnetic reconnection, demonstrating a reasonable match between observations (filled circles) and the model prediction (the black line with a slope of $\sim$6.55). }
\label{fig:mrx_Tratio}
\end{figure}

Furthermore, we can examine the temperature ratio $\Delta T_{\rm i}/\Delta T_{\rm e}$.  Based on Eq.(\ref{eq:formula-mrx}) with $L=2d_{\rm i}$ for ion heating and  $L=2\sqrt{d_{\rm i}d_{\rm e}}$ for electron heating, we obtain
\begin{equation}
    \frac{\Delta T_i}{\Delta T_e} = \left(\frac{m_{\rm i}}{m_e}\right)^{\frac{1}{4}} \sim 6.55
\end{equation}
This does not depend on the reconnection rate $\alpha_{\rm R}$ and could provide a universal upper limit to this heating ratio. Figure \ref{fig:mrx_Tratio} shows the observed $\Delta T_{\rm i}$ versus $\Delta T_{\rm e}$ \mcl{(filled circles) as well as the relation obtained from the derived formulas (thin line). It is evident that the prediction of the model matches well with the observed temperature ratios.}

If we assume $\Delta T_i/\Delta T_e \sim T_i/T_e$, our predicted ratio is within the range of empirically known values of $T_{\rm i}/T_{\rm e} \sim 2-10$ \citep[e.g.][]{SlavinJA_1985_ISEE, BaumjohannW_1989_Average, WangCP_2012_Spatial, WatanabeK_2019_Statistical}. It is interesting that a theoretical study of particle motion in a modeled magnetotail predicts $T_{\rm i}/T_{\rm e} \propto (m_{\rm i}/m_{\rm e})^{1/3}$ \citep{SchriverD_1998_origin}. Also, a more recent theoretical study based on effective ohmic heating predicts a similar scaling as ours, $\Delta T_{\rm i}/\Delta T_{\rm e} = (m_{\rm i}T_{\rm i0}/m_{\rm e}T_{\rm e0})^{1/4} \sim (m_{\rm i}/m_{\rm e})^{1/4}$, where $T_{\rm i0}$ and $T_{\rm e0}$ are the ion and electron temperature in the plasma sheet before reconnection onset \citep{HoshinoM_2018_Energy}.

\section{Summary and Discussion}\label{sec:discuss}

\begin{deluxetable*}{lCCC}[t]
\tablecaption{Scaling of particle heating derived in this study\label{tab:formula}}
\tablehead{
\colhead{} & \colhead{\rule{0pt}{4ex} $ \displaystyle \Delta T_{\rm i}$} & \colhead{$\displaystyle \Delta T_{\rm e}$} & \colhead{$\displaystyle \frac{\Delta T_{\rm i}}{\Delta T_{\rm e}}$}\\[-0.4cm]}
\startdata
Shock & \rule{0pt}{5ex} \displaystyle \quad\quad\frac{4}{3}M_{\rm A}\left(\frac{3}{2}\beta_{\rm i}\right)^{\frac{1}{2}}m_{\rm i}V_{\rm A}^2 & \displaystyle \quad\quad\frac{4}{3}M_{\rm A}\left(\frac{m_{\rm i}}{m_{\rm e}}\right)^{-\frac{1}{4}}m_{\rm i}V_{\rm A}^2 & \displaystyle \quad\quad \left(\frac{m_{\rm i}}{m_{\rm e}}\right)^{\frac{1}{4}} \left(\frac{3}{2}\beta_{\rm i}\right)^{\frac{1}{2}} \\ [0.4cm]
 & \rule{0pt}{5ex} (L=2r_{\rm gi}) & (L=2\sqrt{d_{\rm i}d_{\rm e}}) & \\[0.4cm]\hline
Magnetic Reconnection & \rule{0pt}{5ex} \displaystyle \quad\quad \frac{2}{15}\,\,m_{\rm i}V_{\rm A}^2$ & \displaystyle \quad\quad $\frac{2}{15}\left(\frac{m_{\rm i}}{m_{\rm e}}\right)^{-\frac{1}{4}}m_{\rm i}V_{\rm A}^2 & \displaystyle \quad\quad \left(\frac{m_{\rm i}}{m_{\rm e}}\right)^{\frac{1}{4}} \\ [0.4cm]
 & \rule{0pt}{5ex} (L=2d_{\rm i}) & (L=2\sqrt{d_{\rm i}d_{\rm e}}) & \\[0.4cm]
\enddata
\tablecomments{The governing formula is $\Delta T=(2/3)(V_{\rm in}/V_{\rm A})(L/d_{\rm i})m_{\rm i}V_{\rm A}^2$, as described in Section \ref{sec:model}. The relevant length scales $L$, shown in parentheses, were inferred from comparisons with observations. For reconnection, $L$ is derived with an assumed reconnection rate of $\alpha_R=V_{\rm in}/V_{\rm A}=0.1$.}
\end{deluxetable*}

In summary, starting from the simple expression $\Delta \varepsilon = qVBL$, we showed that the scaling of particle heating for both shocks and magnetic reconnection can be explained by the product of the Alfv\'{e}n Mach number $M_{\rm A}$ (or the reconnection rate $\alpha_{\rm R}$) and the characteristic spatial scale $L/d_{\rm i}$. We found that $L$ is on the order of (1) the ion gyro-radius for ion heating at shocks, (2) the ion inertia length for ion heating in reconnection, and (3) the hybrid inertia length for electron heating in both shocks and reconnection. With these scales, we obtained reduced formulas, as compiled in Table \ref{tab:formula}.
The fact that the spatial scale $L$ is similar for shocks and magnetic reconnection, and that $\Delta T_{\rm i}/\Delta T_{\rm e}$ shows the same dependence on the ion-to-electron mass ratio, suggests  a universality in the physics of particle heating by shocks and magnetic reconnection. However, an epistemological issue remains regarding how these choices of $L$ can be justified. While we do not have clear answers, we will discuss some ideas and issues below.

For ion heating at supercritical shocks, it is well known that energy dissipation (and thus ion heating) is assisted by the reflection and subsequent gyration of a small fraction of incoming ions, resulting in shock drift acceleration \citep[e.g.][]{LeroyMM_1981_Simulation, PaschmannG_1982_Observations}. The distance of travel during the energization is about twice the ion gyro-radius. However, the gyro-radius of such reflected-gyrating ions $r_{\rm gi,in}$ should be based on the incoming, supersonic flow speed, $V_{\rm in}\sin{\theta_{\rm Bn}}$ \citep{GoslingJT_1982_Evidence} and thus very large. In fact, as we examined in Section \ref{sec:shock}, the value of $L$ that fits the data is a factor of $\sim1/4$ smaller than $r_{\rm gi,in}$. \mcl{We found that $L \sim 2 r_{\rm gi}$, where $r_{\rm gi}$ is the ion gyro-radius based on the upstream ion thermal speed, can reproduce the observed values of $\Delta T_{\rm i}$. This result implies that $\Delta T_{\rm i}$ depends on $\beta_{\rm i}$. Interestingly, a recent study of subcritical interplanetary shocks, rather than supercritical bow shocks as in our case, indicates that the width of the ion transition scale in the shock ramp is approximately $3r_{\rm gi}$ \citep{NemecekZ_2013_Ion}. This suggests that  a similar heating process may occur in both supercritical and subcritical shocks}.

It should also be mentioned that the reflected-gyrating ions and transmitted ions would eventually thermalize in the downstream region due to plasma instabilities and that the typical wavelength of the relevant waves are on the order of ion gyro-radius \citep[e.g.][]{WuCS_1984_Microinstabilities}. 

For ion heating in magnetic reconnection, we found that $L=2d_{\rm i}$ reproduces the observed scaling, but $L=2r_{\rm gi}$ can also approximate the temperature increase fairly well. In this regard, a classical, particle description of magnetic reconnection envisions a gyration of ions around the normal ($z$) component of field lines moving in the jet  direction, resulting in an energization by the reconnection electric field in the out-of-plane ($y$) direction \citep[e.g.][]{CowleySWH_1985_Magnetic}. The distance of travel along the $y$-direction is about two times the gyro-radius which corresponds to the diameter of the gyromotion. Therefore, it may not be surprising that $L=2r_{\rm gi}$, where $r_{\rm gi}$ is defined with the thermal velocity in the {\it outflow} region, was able to match the observations. 

For electron heating, we found that using the hybrid scale $2\sqrt{d_{\rm i}d_{\rm e}}$ \mcl{(or alternatively $2\sqrt{r_{\rm gi}r_{\rm ge}}$)} in the expression works well in both shocks and magnetic reconnection, indicating that the heating process involves coupled ion and electron dynamics. In fact, previous studies of quasi-perpendicular shocks have argued that electron heating is associated with the shock potential that develops in the ramp region due to charge separation between ions and electrons \citep[e.g.][]{GoodrichCC_1984_adiabatic}. The width of the ramp region could therefore represent the spatial scale of electron heating and many studies show that the width is comparable to or somewhat smaller than the ion scale \citep[e.g.][and references therein]{JohlanderA_2023_Electron}. For magnetic reconnection, the hybrid gyro-radius $\sqrt{r_{\rm gi}r_{\rm ge}}$ could be a characteristic length scale due to charge separation at the magnetopause current sheet \citep[e.g.][]{CowleySWH_1985_Magnetic} while the hybrid inertial length $\sqrt{d_{\rm i}d_{\rm e}}$ appears in the scaling of the size of the electron diffusion region \citep{NakamuraT_2016_Spatial}. It should also be mentioned that the hybrid scale is a key characteristics of the lower-hybrid (whistler) waves, which play an important role in both physics of shocks and magnetic reconnection.

We anticipate that more sophisticated modeling using simulations, combined with analysis of larger sets of observations, will reveal more detailed insight into the scaling law for particle heating at shocks and in reconnecting current sheets, including more precise meaning of the spatial scale $L$. For shocks, the scaling should be tested for a wider range of parameters such as the shock angle $\theta_{Bn}$  and the plasma $\beta$. \mcl{The shock angle dependence is particularly important as our formula $\varepsilon=qVBL$ assumes a perpendicular orientation of the magnetic field and our analysis was indeed limited to quasi-perpendicular shocks.} For magnetic reconnection, we emphasize again that it is the product $\alpha_{\rm R}L$ that determines the scaling. If $\alpha_{\rm R}$ has a different value, then the value of $L$ would have to be adjusted. For example, if $\alpha_{\rm R}$=0.2, then $L$ would be half of the derived value.  Further dependence on other parameters such as the upstream plasma beta, the inflow-to-outflow density ratio, and the guide field should also be examined. Such detailed studies are left for future work.


We thank Kazuo Makishima and Toshio Terasawa for valuable discussions, particularly regarding the interpretation of $f$ and $L$. 
Some of the shock parameters were obtained from https://sharp.fmi.fi/shock-database/ and developed in the SHARP project, funded by the European Union under grant agreement number 101004131. MO was supported by NASA grants 80NSSC18K1002 and 80NSSC22K0520 at UC Berkeley. M\O{} was supported by NSF grant PHY-2409449 at UC Berkeley.

%






\appendix

\section{Selection and analysis of bow shock crossings}

To investigate particle heating at shocks and how our formula Eq.(\ref{eq:formula-shock}) can apply, we analyzed data from Earth's bow shock obtained by {\it MMS}. First, we examined the publicly-available shock-crossing catalog generated with a machine learning approach \citep{LaltiA_2022_Database}.  Through visual inspection, we selected \mcl{102} `clean' supercritical shock crossings that showed no upstream ions and no signs of partial crossings. \mcl{As a result, our events focus on quasi-perpendicular shocks with a relatively large shock angle, $\theta_{Bn} >60^{\rm o}$. It is worth noting that, when we were searching for the events, we encountered a few cases that did not show energetic ions --  specifically, ions more energetic than the incoming solar wind ions -- in the shock transition layer and the immediate downstream region. Such cases are likely to be subcritical shocks and were therefore excluded from our samples of 102 crossings.}

For the upstream parameters such as the Alfv\'{e}n Mach number $M_{\rm A}$, ion density $N_{\rm i, up}$, and ion temperature $T_{\rm i, up}$, we used the values based on time-shifted data from spacecraft located upstream of {\it MMS} ({\it OMNI} database) \citep{KingJH_2005_Solar}.  These values are already included in the catalog. Throughout this study, $M_{\rm A}$ is defined in the shock-rest, shock normal incidence frame, in which the upstream flow is projected to the shock normal vector $\hat{\mathbf{n}}$. While \cite{LaltiA_2022_Database} used the bow shock model by \cite{FarrisMH_1991_thickness} to derive $\hat{\mathbf{n}}$ and the shock angle $\theta_{\rm Bn}$, we used the model by \cite{PeredoM_1995_Threedimensional}. For the shock propagation speed $V_{\rm shock}$, we used an empirical relation derived from a recent study using the timing method: $V_{\rm shock}=0.04*V_{\rm SW}+1.31$ (km/s), where $V_{SW}$ is the solar wind speed (km/s) \citep{KruparovaO_2019_Statistical}. We confirmed that our results are not sensitive to the estimation of $V_{\rm shock}$. We obtain  essentially the same results even if we did not correct for $V_{\rm shock}$.

\begin{figure*}
\plotone{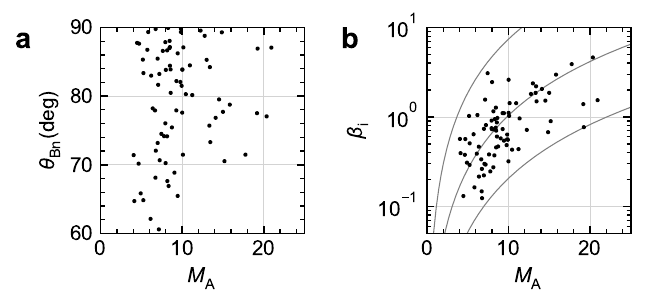}
\caption{The upstream parameter space covered by our statistical study of Earth's bow shock using {\it MMS}.  (a) $M_{\rm A}$ versus $\theta_{\rm Bn}$, (b) $M_{\rm A}$ versus $\beta_{\rm i}$. The curve shows $\beta_{\rm i}=(2/3)\psi^{-2}M_{\rm A}^2$ with  $\psi=$3, 8, and 18.
\label{fig:shock_param}}
\end{figure*}

For the upstream electron temperature $T_{\rm e, up}$, we followed \cite{LaltiA_2022_Database} and used the canonical  value of $T_{\rm e, up} = 12.06$ eV. We confirmed that using $T_{\rm e, up}$ measured by {\it MMS} do not change the main conclusion of this paper. For $N_{\rm i,up}$, we found that the values measured by {\it MMS} and those of {\it OMNI} are often different, consistent with earlier findings \citep{RobertsOW_2021_Study}, leading to some outliers in our plots.  While our main conclusion of this paper are not affected by those outliers, we removed such cases by requiring that that the difference in $N_{\rm i,up}$  between {\it MMS} and {\it OMNI} is less than 20\%. This filtering resulted in a reduction of the total number of events from \mcl{102 to 80}.  For the key upstream parameters ($M_{\rm A}$, $\theta_{\rm Bn}$, $\beta_{\rm i}$), the range of values in our dataset are $M_{\rm A} \sim 4 - 20$, $\theta_{Bn} \sim 60^{\rm o} - 90^{\rm o}$, and $\beta_{\rm i} \sim 0.1 - 6$, as shown in Figure \ref{fig:shock_param}. It is evident that there is a rough correlation between $M_{\rm A}$ and $\beta_{\rm i}$ due to the intrinsic relation of $\beta_{\rm i}=(2/3)\psi^{-2}M_{\rm A}^2$, where $\psi=V_{\rm in}/V_{\rm thi}$. The solid curves in Figure \ref{fig:shock_param}b shows the case for $\psi=$3, 8, and 18.

\mcl{A list of all 80 shock crossing events is provided in a separate file in the Comma-Separated Values (CSV) format. The list includes the upstream and downstream time intervals and all physical parameters that are necessary to reproduce Figures \ref{fig:shock}-\ref{fig:shock-beta}. The columns in the CSV  file are briefly explained in Table \ref{list_shock}.}

\section{List of magnetic reconnection events}

\mcl{For our statistical study of magnetic reconnection, we used dataset from both magnetopause and magnetotail. The magnetopause events were obtained by THEMIS and reported by \cite{PhanTD_2013_Electron, PhanTD_2014_Ion}, whereas the magnetotail events were obtained by MMS and reported by \cite{OierosetM_2023_Scaling, OierosetM_2024_Scaling}. }

\mcl{A list of all 103 magnetopause events is provided in a separate file in CSV format. 87 of them are used for ion heating study and 79 of them were used for electron heating study. These events were used in Figures \ref{fig:mrx} and \ref{fig:mrx-rgi-di}. There are 65 events that provide both ion and electron temperature. They were used in producing Figure \ref{fig:mrx_Tratio}. The columns in the CSV file are briefly explained in Table \ref{list_magnetopause}. The Alfv\'{e}n speed in $m_{\rm i}V_{\rm A}^2$ has been calculated with the asymmetry in the inflow direction  taken into account. For more details of how these data were obtained, readers are referred to \cite{PhanTD_2013_Electron, PhanTD_2014_Ion}. }

\mcl{While a complete list of all 20 magnetotail events has already been published as a fully readable table in \cite{OierosetM_2023_Scaling, OierosetM_2024_Scaling}, we have also provided a separate file in CSV format for completeness. The columns in the CSV file are briefly explained in Table \ref{list_magnetotail}.}


\begin{deluxetable}{llCl}[t]
\tablecaption{Descriptive table of the shock crossing events \label{list_shock}}
\tablehead{\colhead{Column Name} & \colhead{Description} & \colhead{Unit} & \colhead{Example Data}}
\startdata
{\fontfamily{qcr}\selectfont up\_tstart } & upstream start time                                                 & \cdots           & {\fontfamily{qcr}\selectfont 2015-11-13/12:55:00} \\
{\fontfamily{qcr}\selectfont up\_tend   } & upstream end time                                                   & \cdots           & {\fontfamily{qcr}\selectfont 2015-11-13/12:56:00} \\
{\fontfamily{qcr}\selectfont dn\_tstart } & downstream start time                                               & \cdots           & {\fontfamily{qcr}\selectfont 2015-11-13/12:52:00} \\
{\fontfamily{qcr}\selectfont dn\_tend   } & downstream end time                                                 & \cdots           & {\fontfamily{qcr}\selectfont 2015-11-13/12:53:00} \\
{\fontfamily{qcr}\selectfont Man        } & Alfven Mach number in the normal incidence frame $M_{\rm A}$        & \cdots           & {\fontfamily{qcr}\selectfont 11.235}  \\
{\fontfamily{qcr}\selectfont tBn        } & shock angle $\theta_{\rm Bn}$                                       & ^\circ           & {\fontfamily{qcr}\selectfont 77.369}  \\
{\fontfamily{qcr}\selectfont up\_Vin    } & upstream bulk flow speed in the normal incidence frame $V_{\rm in}$ & {\rm km\,s}^{-1} & {\fontfamily{qcr}\selectfont 388.266} \\
{\fontfamily{qcr}\selectfont up\_B      } & upstream magnetic field strength $B_{\rm up}$                       & {\rm nT}         & {\fontfamily{qcr}\selectfont 8.167}   \\
{\fontfamily{qcr}\selectfont up\_Ni     } & upstream ion density $N_{\rm i, up}$                                & {\rm cm}^{-3}    & {\fontfamily{qcr}\selectfont 26.542}  \\
{\fontfamily{qcr}\selectfont up\_Ti     } & upstream ion temperature  $T_{\rm i,up}$                            & {\rm eV}         & {\fontfamily{qcr}\selectfont 2.73}    \\
{\fontfamily{qcr}\selectfont up\_Te     } & upstream electron temperature  $T_{\rm e,up}$                       & {\rm eV}         & {\fontfamily{qcr}\selectfont 12.06} \\
{\fontfamily{qcr}\selectfont dn\_Ti     } & downstream ion temperature  $T_{\rm i,dn}$                          & {\rm eV}         & {\fontfamily{qcr}\selectfont 255.796} \\
{\fontfamily{qcr}\selectfont dn\_Te     } & downstream electron temperature $T_{\rm e}$                         & {\rm eV}         & {\fontfamily{qcr}\selectfont 42.33} 
\enddata
\tablecomments{This data file contains 13 columns, each of which is described in the table along with the relevant units and an example data value. The data file is available in its entirety in the machine-readable CSV format in the online article. }
\end{deluxetable}

\begin{deluxetable}{llCl}[t]
\tablecaption{Descriptive table of the list of magnetopause events\label{list_magnetopause}}
\tablehead{\colhead{Column Name} & \colhead{Description} & \colhead{Unit} & \colhead{Example Data}}
\startdata
{\fontfamily{qcr}\selectfont sh\_Te\_time\_start}& Magnetosheath start time for electron temperature $T_{\rm e}$ & \cdots   & {\fontfamily{qcr}\selectfont 2008-06-26/19:57:22 }\\
{\fontfamily{qcr}\selectfont sh\_Te\_time\_end  }& Magnetosheath end   time for electron temperature $T_{\rm e}$ & \cdots   & {\fontfamily{qcr}\selectfont 2008-06-26/19:57:45 }\\
{\fontfamily{qcr}\selectfont sp\_Te\_time\_start}& Magnetosphere start time for electron temperature $T_{\rm e}$ & \cdots   & {\fontfamily{qcr}\selectfont 2008-06-26/19:55:36 }\\
{\fontfamily{qcr}\selectfont sp\_Te\_time\_end  }& Magnetosphere end   time for electron temperature $T_{\rm e}$ & \cdots   & {\fontfamily{qcr}\selectfont 2008-06-26/19:55:44 }\\
{\fontfamily{qcr}\selectfont sh\_Ti\_time\_start}& Magnetosheath start time for ion temperature $T_{\rm i}$      & \cdots   & {\fontfamily{qcr}\selectfont NaN                 }\\
{\fontfamily{qcr}\selectfont sh\_Ti\_time\_end  }& Magnetosheath end   time for ion temperature $T_{\rm i}$      & \cdots   & {\fontfamily{qcr}\selectfont NaN                 }\\
{\fontfamily{qcr}\selectfont sp\_Ti\_time\_start}& Magnetosphere start time for ion temperature $T_{\rm i}$      & \cdots   & {\fontfamily{qcr}\selectfont NaN                 }\\
{\fontfamily{qcr}\selectfont sp\_Ti\_time\_end  }& Magnetosphere end   time for ion temperature $T_{\rm i}$      & \cdots   & {\fontfamily{qcr}\selectfont NaN                 }\\
{\fontfamily{qcr}\selectfont dTe                }& Electron temperature difference $dT_{\rm e}$                  & \rm{eV} & {\fontfamily{qcr}\selectfont 22.541756           }\\
{\fontfamily{qcr}\selectfont Te\_in             }& Electron temperature in the inflow region $T_{\rm e,in}$      & \rm{eV} & {\fontfamily{qcr}\selectfont 43.650467           }\\
{\fontfamily{qcr}\selectfont Te\_out            }& Electron temperature in the outflow region $T_{\rm e,out}$    & \rm{eV} & {\fontfamily{qcr}\selectfont 66.19222            }\\
{\fontfamily{qcr}\selectfont dTi                }& Ion temperature difference $dT_{\rm i}$                       & \rm{eV} & {\fontfamily{qcr}\selectfont NaN                 }\\
{\fontfamily{qcr}\selectfont Ti\_in             }& Ion temperature in the inflow region $T_{\rm i,in}$           & \rm{eV} & {\fontfamily{qcr}\selectfont NaN                 }\\
{\fontfamily{qcr}\selectfont Ti\_out            }& Ion temperature in the outflow region $T_{\rm i,out}$         & \rm{eV} & {\fontfamily{qcr}\selectfont NaN                 }\\
{\fontfamily{qcr}\selectfont miVa2              }& $m_{\rm i}V_{\rm A}^2$                                        & \rm{eV} & {\fontfamily{qcr}\selectfont 1604.7769           }
\enddata
\tablecomments{This data file contains 16 columns, each of which is described in the table along with the relevant units and an example data value. The data file is available in its entirety in the machine-readable CSV format in the online article.  }
\end{deluxetable}

\begin{deluxetable}{llCl}[t]
\tablecaption{Descriptive table of the list of magnetotail events\label{list_magnetotail}}
\tablehead{\colhead{Column Name} & \colhead{Description} & \colhead{Unit} & \colhead{Example Data}}
\startdata
{\fontfamily{qcr}\selectfont tstart\_in  }& Inflow start time                            & \cdots  & {\fontfamily{qcr}\selectfont 2017-07-26/13:00:16} \\
{\fontfamily{qcr}\selectfont tend\_in    }& Inflow end time                              & \cdots  & {\fontfamily{qcr}\selectfont 2017-07-26/13:02:00} \\
{\fontfamily{qcr}\selectfont tstart\_out }& Outflow start time                           & \cdots  & {\fontfamily{qcr}\selectfont 2017-07-26/13:05:22} \\
{\fontfamily{qcr}\selectfont tend\_out   }& Outflow end time                             & \cdots  & {\fontfamily{qcr}\selectfont 2017-07-26/13:06:58} \\
{\fontfamily{qcr}\selectfont  Ti\_in     }& Inflow ion temperature  $T_{\rm i,in}$       & \rm{eV} & {\fontfamily{qcr}\selectfont 7.018} \\
{\fontfamily{qcr}\selectfont  Te\_in     }& Inflow electron temperature  $T_{\rm e,in}$  & \rm{eV} & {\fontfamily{qcr}\selectfont 1.678} \\
{\fontfamily{qcr}\selectfont  Ti\_out    }& Outflow ion temperature  $T_{\rm i,out}$     & \rm{eV} & {\fontfamily{qcr}\selectfont 9.620} \\
{\fontfamily{qcr}\selectfont  Te\_out    }& Outflow electron temperature $T_{\rm e,out}$ & \rm{eV} & {\fontfamily{qcr}\selectfont 2.042} \\
{\fontfamily{qcr}\selectfont  miVa2       }& $m_{\rm i}V_{\rm A}^2$                       & \rm{eV} & {\fontfamily{qcr}\selectfont 14.34}
\enddata
\tablecomments{This data file contains 9 columns, each of which is described in the table along with the relevant units and an example data value. The data file is available in its entirety in the machine-readable CSV format in the online article. }
\end{deluxetable}

\clearpage
\bibliography{references}{} 

\begin{thebibliography}{}
\expandafter\ifx\csname natexlab\endcsname\relax\def\natexlab#1{#1}\fi
\providecommand{\url}[1]{\href{#1}{#1}}
\providecommand{\dodoi}[1]{doi:~\href{http://doi.org/#1}{\nolinkurl{#1}}}
\providecommand{\doeprint}[1]{\href{http://ascl.net/#1}{\nolinkurl{http://ascl.net/#1}}}
\providecommand{\doarXiv}[1]{\href{https://arxiv.org/abs/#1}{\nolinkurl{https://arxiv.org/abs/#1}}}

\bibitem[{Angelopoulos(2008)}]{AngelopoulosV_2008_THEMIS}
Angelopoulos, V. 2008, Space Science Reviews, 141, 5,
  \dodoi{10.1007/s11214-008-9336-1}

\bibitem[{Aschwanden(2005)}]{AschwandenMJ_2005_Physics}
Aschwanden, M.~J. 2005, Physics of the {Solar} {Corona} (Springer Berlin
  Heidelberg)

\bibitem[{Baumjohann {et~al.}(1989)Baumjohann, Paschmann, \&
  Cattell}]{BaumjohannW_1989_Average}
Baumjohann, W., Paschmann, G., \& Cattell, C.~A. 1989, Journal of Geophysical
  Research-Space Physics, 94, 6597, \dodoi{DOI 10.1029/JA094iA06p06597}

\bibitem[{Birn {et~al.}(2001)Birn, Drake, Shay, Rogers, Denton, Hesse,
  Kuznetsova, Ma, Bhattacharjee, Otto, \& Pritchett}]{BirnJ_2001_Geospace}
Birn, J., Drake, J.~F., Shay, M.~A., {et~al.} 2001, Journal of Geophysical
  Research-Space Physics, 106, 3715, \dodoi{DOI 10.1029/1999ja900449}

\bibitem[{Bohdan {et~al.}(2020)Bohdan, Pohl, Niemiec, Morris, Matsumoto, Amano,
  \& Hoshino}]{BohdanA_2020_Kinetic}
Bohdan, A., Pohl, M., Niemiec, J., {et~al.} 2020, The Astrophysical Journal,
  904, \dodoi{10.3847/1538-4357/abbc19}

\bibitem[{Burch {et~al.}(2016)Burch, Moore, Torbert, \&
  Giles}]{BurchJL_2016_Magnetospheric}
Burch, J.~L., Moore, T.~E., Torbert, R.~B., \& Giles, B.~L. 2016, Space Science
  Reviews, 199, 5, \dodoi{10.1007/s11214-015-0164-9}

\bibitem[{Cowley(1985)}]{CowleySWH_1985_Magnetic}
Cowley, S. W.~H. 1985, in Solar {System} {Magnetic} {Fields} (Holland: D.
  Reidel Publishing Company), 121--155

\bibitem[{Drake {et~al.}(2009)Drake, Swisdak, Phan, Cassak, Shay, Lepri, Lin,
  Quataert, \& Zurbuchen}]{DrakeJF_2009_Ion}
Drake, J.~F., Swisdak, M., Phan, T.~D., {et~al.} 2009, Journal of Geophysical
  Research: Space Physics, 114, \dodoi{10.1029/2008ja013701}

\bibitem[{Eastwood {et~al.}(2013)Eastwood, Phan, Drake, Shay, Borg, Lavraud, \&
  Taylor}]{EastwoodJP_2013_Energy}
Eastwood, J.~P., Phan, T.~D., Drake, J.~F., {et~al.} 2013, Phys Rev Lett, 110,
  225001, \dodoi{10.1103/PhysRevLett.110.225001}

\bibitem[{Egedal {et~al.}(2013)Egedal, Le, \& Daughton}]{EgedalJ_2013_review}
Egedal, J., Le, A., \& Daughton, W. 2013, Physics of Plasmas, 20, 061201,
  \dodoi{10.1063/1.4811092}

\bibitem[{Egedal {et~al.}(2005)Egedal, Oieroset, Fox, \&
  Lin}]{EgedalJ_2005_situ}
Egedal, J., Oieroset, M., Fox, W., \& Lin, R.~P. 2005, Phys Rev Lett, 94,
  025006, \dodoi{10.1103/PhysRevLett.94.025006}

\bibitem[{Farris {et~al.}(1991)Farris, Petrinec, \&
  Russell}]{FarrisMH_1991_thickness}
Farris, M.~H., Petrinec, S.~M., \& Russell, C.~T. 1991, Geophysical Research
  Letters, 18, 1821, \dodoi{10.1029/91gl02090}

\bibitem[{Genestreti {et~al.}(2018)Genestreti, Nakamura, Nakamura, Denton,
  Torbert, Burch, Plaschke, Fuselier, Ergun, Giles, \&
  Russell}]{GenestretiKJ_2018_How}
Genestreti, K.~J., Nakamura, T. K.~M., Nakamura, R., {et~al.} 2018, Journal of
  Geophysical Research (Space Physics), 123, 9130, \dodoi{10.1029/2018ja025711}

\bibitem[{Ghavamian {et~al.}(2013)Ghavamian, Schwartz, Mitchell, Masters, \&
  Laming}]{GhavamianP_2013_ElectronIon}
Ghavamian, P., Schwartz, S.~J., Mitchell, J., Masters, A., \& Laming, J.~M.
  2013, Space Science Reviews, 178, 633, \dodoi{10.1007/s11214-013-9999-0}

\bibitem[{Goodrich \& Scudder(1984)}]{GoodrichCC_1984_adiabatic}
Goodrich, C.~C., \& Scudder, J.~D. 1984, Journal of Geophysical Research: Space
  Physics, 89, 6654, \dodoi{10.1029/JA089iA08p06654}

\bibitem[{Gosling {et~al.}(1982)Gosling, Thomsen, Bame, Feldman, Paschmann, \&
  Sckopke}]{GoslingJT_1982_Evidence}
Gosling, J.~T., Thomsen, M.~F., Bame, S.~J., {et~al.} 1982, Geophysical
  Research Letters, 9, 1333, \dodoi{10.1029/GL009i012p01333}

\bibitem[{Haggerty {et~al.}(2018)Haggerty, Shay, Chasapis, Phan, Drake,
  Malakit, Cassak, \& Kieokaew}]{HaggertyCC_2018_reduction}
Haggerty, C.~C., Shay, M.~A., Chasapis, A., {et~al.} 2018, Physics of Plasmas,
  25, \dodoi{10.1063/1.5050530}

\bibitem[{Haggerty {et~al.}(2015)Haggerty, Shay, Drake, Phan, \&
  McHugh}]{HaggertyCC_2015_competition}
Haggerty, C.~C., Shay, M.~A., Drake, J.~F., Phan, T.~D., \& McHugh, C.~T. 2015,
  Geophysical Research Letters, 42, 9657, \dodoi{10.1002/2015gl065961}

\bibitem[{Hillas(1984)}]{HillasM_1984_Origin}
Hillas, a.~M. 1984, Annual Review of Astronomy and Astrophysics, 22, 425,
  \dodoi{10.1146/annurev.aa.22.090184.002233}

\bibitem[{Hoshino(2018)}]{HoshinoM_2018_Energy}
Hoshino, M. 2018, The Astrophysical Journal Letters, 868,
  \dodoi{10.3847/2041-8213/aaef3a}

\bibitem[{Hull {et~al.}(2000)Hull, Scudder, Fitzenreiter, Ogilvie, Newbury, \&
  Russell}]{HullAJ_2000_Electron}
Hull, A.~J., Scudder, J.~D., Fitzenreiter, R.~J., {et~al.} 2000, Journal of
  Geophysical Research: Space Physics, 105, 20957, \dodoi{10.1029/2000ja900049}

\bibitem[{Johlander {et~al.}(2023)Johlander, Khotyaintsev, Dimmock, Graham, \&
  Lalti}]{JohlanderA_2023_Electron}
Johlander, A., Khotyaintsev, Y.~V., Dimmock, A.~P., Graham, D.~B., \& Lalti, A.
  2023, Geophysical Research Letters, 50, \dodoi{10.1029/2022gl100400}

\bibitem[{King \& Papitashvili(2005)}]{KingJH_2005_Solar}
King, J.~H., \& Papitashvili, N.~E. 2005, Journal of Geophysical Research:
  Space Physics, 110, \dodoi{10.1029/2004ja010649}

\bibitem[{Kivelson \& Russell(1995)}]{KivelsonMG_1995_Introduction}
Kivelson, M.~G., \& Russell, C.~T. 1995, Introduction to {Space} {Physics}
  (Cambridge University Press)

\bibitem[{Kruparova {et~al.}(2019)Kruparova, Krupar, Safránková, Nemecek,
  Maksimovic, Santolik, Soucek, Nemec, \& Merka}]{KruparovaO_2019_Statistical}
Kruparova, O., Krupar, V., Safránková, J., {et~al.} 2019, Journal of
  Geophysical Research-Space Physics, 124, 1539, \dodoi{10.1029/2018ja026272}

\bibitem[{Lalti {et~al.}(2022)Lalti, Khotyaintsev, Dimmock, Johlander, Graham,
  \& Olshevsky}]{LaltiA_2022_Database}
Lalti, A., Khotyaintsev, Y.~V., Dimmock, A.~P., {et~al.} 2022, Journal of
  Geophysical Research: Space Physics, 127, \dodoi{10.1029/2022ja030454}

\bibitem[{Leroy {et~al.}(1981)Leroy, Goodrich, Winske, Wu, \&
  Papadopoulos}]{LeroyMM_1981_Simulation}
Leroy, M.~M., Goodrich, C.~C., Winske, D., Wu, C.~S., \& Papadopoulos, K. 1981,
  Geophysical Research Letters, 8, 1269, \dodoi{10.1029/gl008i012p01269}

\bibitem[{Lin \& Lee(1994)}]{LinY_1994_Structure}
Lin, Y., \& Lee, L.~C. 1994, Space Science Reviews, 65, 59,
  \dodoi{10.1007/bf00749762}

\bibitem[{Liu {et~al.}(2022)Liu, Cassak, Li, Hesse, Lin, \&
  Genestreti}]{LiuYH_2022_Firstprinciples}
Liu, Y.-H., Cassak, P., Li, X., {et~al.} 2022, Communications Physics, 5,
  \dodoi{10.1038/s42005-022-00854-x}

\bibitem[{Liu {et~al.}(2017)Liu, Hesse, Guo, Daughton, Li, Cassak, \&
  Shay}]{LiuYH_2017_Why}
Liu, Y.-H., Hesse, M., Guo, F., {et~al.} 2017, Physical Review Letters, 118,
  \dodoi{10.1103/physrevlett.118.085101}

\bibitem[{Livesey {et~al.}(1984)Livesey, Russell, \&
  Kennel}]{LiveseyWA_1984_comparison}
Livesey, W.~A., Russell, C.~T., \& Kennel, C.~F. 1984, Journal of Geophysical
  Research: Space Physics, 89, 6824, \dodoi{10.1029/JA089iA08p06824}

\bibitem[{Longair(2011)}]{LongairMS_2011_High}
Longair, M.~S. 2011, High {Energy} {Astrophysics}, 3rd edn. (United Kingdom:
  Cambridge University Press)

\bibitem[{Matthaeus {et~al.}(1984)Matthaeus, Ambrosiano, \&
  Goldstein}]{MatthaeusWH_1984_Particle}
Matthaeus, W.~H., Ambrosiano, J.~J., \& Goldstein, M.~L. 1984, Physical Review
  Letters, 53, 1449, \dodoi{10.1103/PhysRevLett.53.1449}

\bibitem[{Nakamura {et~al.}(2016)Nakamura, Nakamura, \&
  Haseagwa}]{NakamuraT_2016_Spatial}
Nakamura, T., Nakamura, R., \& Haseagwa, H. 2016, Annales Geophysicae, 34, 357,
  \dodoi{10.5194/angeo-34-357-2016}

\bibitem[{Nakamura {et~al.}(2018)Nakamura, Genestreti, Liu, Nakamura, Teh,
  Hasegawa, Daughton, Hesse, Torbert, Burch, \&
  Giles}]{NakamuraTKM_2018_Measurement}
Nakamura, T. K.~M., Genestreti, K.~J., Liu, Y.-H., {et~al.} 2018, Journal of
  Geophysical Research (Space Physics), 123, 9150, \dodoi{10.1029/2018ja025713}

\bibitem[{Němeček {et~al.}(2013)Němeček, Šafránková, Goncharov, Přech,
  \& Zastenker}]{NemecekZ_2013_Ion}
Němeček, Z., Šafránková, J., Goncharov, O., Přech, L., \& Zastenker,
  G.~N. 2013, Geophysical Research Letters, 40, 4133, \dodoi{10.1002/grl.50814}

\bibitem[{Oka {et~al.}(2025)Oka, Makishima, \& Terasawa}]{OkaM_2025_Maximum}
Oka, M., Makishima, K., \& Terasawa, T. 2025, The Astrophysical Journal, 979,
  161, \dodoi{10.3847/1538-4357/ad9916}

\bibitem[{Oka {et~al.}(2022)Oka, Phan, Øieroset, Turner, Drake, Li, Fuselier,
  Gershman, Giles, Ergun, Torbert, Wei, Strangeway, Russell, \&
  Burch}]{OkaM_2022_Electron}
Oka, M., Phan, T.~D., Øieroset, M., {et~al.} 2022, Physics of Plasmas, 29,
  052904, \dodoi{10.1063/5.0085647}

\bibitem[{Paschmann {et~al.}(1982)Paschmann, Sckopke, Bame, \&
  Gosling}]{PaschmannG_1982_Observations}
Paschmann, G., Sckopke, N., Bame, S.~J., \& Gosling, J.~T. 1982, Geophysical
  Research Letters, 9, 881, \dodoi{10.1029/GL009i008p00881}

\bibitem[{Peredo {et~al.}(1995)Peredo, Slavin, Mazur, \&
  Curtis}]{PeredoM_1995_Threedimensional}
Peredo, M., Slavin, J.~a., Mazur, E., \& Curtis, S.~a. 1995, Journal of
  Geophysical Research: Space Physics, 100, 7907, \dodoi{10.1029/94JA02545}

\bibitem[{Phan {et~al.}(2013)Phan, Shay, Gosling, Fujimoto, Drake, Paschmann,
  Oieroset, Eastwood, \& Angelopoulos}]{PhanTD_2013_Electron}
Phan, T.~D., Shay, M.~A., Gosling, J.~T., {et~al.} 2013, Geophysical Research
  Letters, 40, 4475, \dodoi{10.1002/grl.50917}

\bibitem[{Phan {et~al.}(2014)Phan, Drake, Shay, Gosling, Paschmann, Eastwood,
  Oieroset, Fujimoto, \& Angelopoulos}]{PhanTD_2014_Ion}
Phan, T.~D., Drake, J.~F., Shay, M.~A., {et~al.} 2014, Geophysical Research
  Letters, 41, 7002, \dodoi{10.1002/2014gl061547}

\bibitem[{Priest \& Forbes(2000)}]{PriestE_2000_Magnetic}
Priest, E., \& Forbes, T. 2000, Magnetic {Reconnection} (Cambridge University
  Press)

\bibitem[{Raymond {et~al.}(2023)Raymond, Ghavamian, Bohdan, Ryu, Niemiec,
  Sironi, Tran, Amato, Hoshino, Pohl, Amano, \&
  Fiuza}]{RaymondJC_2023_Electron}
Raymond, J.~C., Ghavamian, P., Bohdan, A., {et~al.} 2023, The Astrophysical
  Journal, 949, \dodoi{10.3847/1538-4357/acc528}

\bibitem[{Roberts {et~al.}(2021)Roberts, Nakamura, Coffey, Gershman, Volwerk,
  Varsani, Giles, Dorelli, \& Pollock}]{RobertsOW_2021_Study}
Roberts, O.~W., Nakamura, R., Coffey, V.~N., {et~al.} 2021, Journal of
  Geophysical Research: Space Physics, 126, e2021JA029784,
  \dodoi{10.1029/2021ja029784}

\bibitem[{Schriver {et~al.}(1998)Schriver, Ashour‐Abdalla, \&
  Richard}]{SchriverD_1998_origin}
Schriver, D., Ashour‐Abdalla, M., \& Richard, R.~L. 1998, Journal of
  Geophysical Research: Space Physics, 103, 14879, \dodoi{10.1029/98ja00017}

\bibitem[{Schwartz {et~al.}(1988)Schwartz, Thomsen, Bame, \&
  Stansberry}]{SchwartzSJ_1988_Electron}
Schwartz, S.~J., Thomsen, M.~F., Bame, S.~J., \& Stansberry, J. 1988, Journal
  of Geophysical Research: Space Physics, 93, 12923,
  \dodoi{10.1029/JA093iA11p12923}

\bibitem[{Shay {et~al.}(1999)Shay, Drake, Rogers, \&
  Denton}]{ShayMA_1999_scaling}
Shay, M.~A., Drake, J.~F., Rogers, B.~N., \& Denton, R.~E. 1999, Geophys Res
  Lett, 26, 2163, \dodoi{Doi 10.1029/1999gl900481}

\bibitem[{Shay {et~al.}(2014)Shay, Haggerty, Phan, Drake, Cassak, Wu, Oieroset,
  Swisdak, \& Malakit}]{ShayMA_2014_Electron}
Shay, M.~A., Haggerty, C.~C., Phan, T.~D., {et~al.} 2014, Physics of Plasmas,
  21, 122902, \dodoi{10.1063/1.4904203}

\bibitem[{Slavin {et~al.}(1985)Slavin, Smith, Sibeck, Baker, Zwickl, \&
  Akasofu}]{SlavinJA_1985_ISEE}
Slavin, J.~A., Smith, E.~J., Sibeck, D.~G., {et~al.} 1985, Journal of
  Geophysical Research: Space Physics, 90, 10875,
  \dodoi{10.1029/JA090iA11p10875}

\bibitem[{Thomsen {et~al.}(1987)Thomsen, Mellott, Stansberry, Bame, Gosling, \&
  Russell}]{ThomsenMF_1987_Strong}
Thomsen, M.~F., Mellott, M.~M., Stansberry, J.~A., {et~al.} 1987, Journal of
  Geophysical Research, 92, 10119, \dodoi{10.1029/JA092iA09p10119}

\bibitem[{Tidman \& Krall(1971)}]{TidmanDA_1971_Shock}
Tidman, D.~A., \& Krall, N.~A. 1971, Shock {Waves} in {Collisionless} {Plasmas}
  (Wiley-Interscience)

\bibitem[{Tran \& Sironi(2020)}]{TranA_2020_Electron}
Tran, A., \& Sironi, L. 2020, The Astrophysical Journal Letters, 900,
  \dodoi{10.3847/2041-8213/abb19c}

\bibitem[{Tsurutani \& Stone(1985)}]{TsurutaniBT_1985_Collisionless}
Tsurutani, B.~T., \& Stone, R.~G. 1985, Geophysical {Monograph} {Series},
  Vol.~35, Collisionless {Shocks} in the {Heliosphere}: {Reviews} of {Current}
  {Research} (American Geophysical Union)

\bibitem[{Vink {et~al.}(2015)Vink, Broersen, Bykov, \&
  Gabici}]{VinkJ_2015_electronion}
Vink, J., Broersen, S., Bykov, A., \& Gabici, S. 2015, Astronomy
  {\textbackslash}\& Astrophysics, 579, \dodoi{10.1051/0004-6361/201424612}

\bibitem[{Wang {et~al.}(2012)Wang, Gkioulidou, Lyons, \&
  Angelopoulos}]{WangCP_2012_Spatial}
Wang, C.-P., Gkioulidou, M., Lyons, L.~R., \& Angelopoulos, V. 2012, Journal of
  Geophysical Research: Space Physics, 117, \dodoi{10.1029/2012ja017658}

\bibitem[{Watanabe {et~al.}(2019)Watanabe, Keika, Hoshino, Kitamura, Saito,
  Giles, \& Paterson}]{WatanabeK_2019_Statistical}
Watanabe, K., Keika, K., Hoshino, M., {et~al.} 2019, Geophysical Research
  Letters, 46, 14223, \dodoi{10.1029/2019gl084837}

\bibitem[{Wu {et~al.}(1984)Wu, Winske, Zhou, Tsai, Rodriguez, Tanaka,
  Papadopoulos, Akimoto, Lin, Leroy, \&
  Goodrich}]{WuCS_1984_Microinstabilities}
Wu, C.~S., Winske, D., Zhou, Y.~M., {et~al.} 1984, Space Science Reviews, 37,
  63, \dodoi{10.1007/bf00213958}

\bibitem[{Zweibel \& Yamada(2009)}]{ZweibelEG_2009_Magnetic}
Zweibel, E.~G., \& Yamada, M. 2009, Annual Review of Astronomy and
  Astrophysics, 47, 291, \dodoi{10.1146/annurev-astro-082708-101726}

\bibitem[{Øieroset {et~al.}(2023)Øieroset, Phan, Oka, Drake, Fuselier,
  Gershman, Maheshwari, Giles, Zhang, Guo, Burch, Torbert, \&
  Strangeway}]{OierosetM_2023_Scaling}
Øieroset, M., Phan, T.~D., Oka, M., {et~al.} 2023, The Astrophysical Journal,
  954, \dodoi{10.3847/1538-4357/acdf44}

\bibitem[{Øieroset {et~al.}(2024)Øieroset, Phan, Drake, Starkey, Fuselier,
  Cohen, Haggerty, Shay, Oka, Gershman, Maheshwari, Burch, Torbert, \&
  Strangeway}]{OierosetM_2024_Scaling}
Øieroset, M., Phan, T.~D., Drake, J.~F., {et~al.} 2024, The Astrophysical
  Journal, 971, \dodoi{10.3847/1538-4357/ad6151}

\end{thebibliography}
\bibliographystyle{aasjournal}



\end{document}